\documentclass{raa}           
\usepackage{graphicx,times}
\usepackage{natbib}
\usepackage{lscape}
\usepackage{longtable}
\usepackage{amssymb,amsmath}
\bibpunct{(}{)}{;}{a}{}{,}

\usepackage[a4paper=true,pagebackref=true]{hyperref}
\hypersetup{pdftitle = globalphot, pdfauthor = My name, pdfsubject= The subject, pdfkeywords = keyword1 keyword2 keyword3} 
\hypersetup{colorlinks = true, linkcolor = green, anchorcolor = red, citecolor = blue, filecolor = red, pagecolor = red, urlcolor = red}

\begin{document}
\title{Global photometric analysis of galactic HII regions$^*$\footnotetext{\small $*$ Supported by the Program 7 of the Presidium of the RAS.}}

\volnopage{ {\bf 2017} Vol.\ {\bf X} No. {\bf XX}, 000--000}
\setcounter{page}{1}

\author{Anastasiia Topchieva\inst{1}, Dmitri Wiebe\inst{1}, Maria S. Kirsanova\inst{1,2}}

\institute{Institute of Astronomy, Russian Academy of Sciences, Moscow 119017, Russia; {\it stasyat@inasan.ru}\\
\and
            Ural Federal University, 19 Mira Str., Ekaterinburg, Russia\\
}
\vs \no 
         {\small Received 2012 June 12; accepted 2012 July 27}

\abstract{Total infrared fluxes are estimated for 99 HII regions around massive stars. The following wavebands have been used for the analysis: 8 and 24 $\mu$m, based on data from Spitzer space telescope (IRAC and MIPS, respectively); 70, 160, 250, 350, and 500 $\mu$m, based on data from Herschel Space Observatory (PACS and SPIRE). The estimated fluxes are used to evaluate the mass fraction of  polycyclic aromatic hydrocarbons ($q_{\rm PAH}$) and the intensity of the ultraviolet emission in the studied objects. It is shown that the PAH mass fraction, $q_{\rm PAH}$, is much lower in these objects than the average Galactic value, implying effective destruction of aromatic particles in HII regions. Estimated radiation field intensities ($U$) are close to those derived for extragalactic HII complexes. Color indices [$F_{24}$/$F_{8}$], [$F_{70}$/$F_{24}$], [$F_{160}$/$F_{24}$], [$F_{160}$/$F_{70}$] are compared to criteria proposed to distinguish between regions of ionized hydrogen and planetary nebulae. Also, we relate our results to analogous color indices for extragalactic complexes of ionized hydrogen.\
}

   \authorrunning{A. Topchieva et al. }
   \titlerunning{Global photometry of HII regions}
   \maketitle

\section{Introduction}
\label{sect:intro}

The amount of new data on the infrared (IR) radiation in our Galaxy grows steadily. Thanks to results, which have been obtained with Spitzer space telescope, we now have the opportunity to study objects, which had been known previously as ring nebulae and are now widely referred to as IR bubbles \citep{2006ApJ...649..759C, 2007ApJ...670..428C}. Their formation is presumably related to the action of massive hot stars on the interstellar material \citep{1988ApJ...329L..93V}. Specifically, it is believed that a bubble appears around an O-B type star, which ionizes surrounding gas and forms an expanding shell due to hot gas pressure and/or powerful stellar wind. 

Observations tend to support this picture. \cite{2010A&A...523A...6D} classified 86\% objects from the \cite{2006ApJ...649..759C} catalogue as HII regions. \cite{2014ApJS..212....1A} created a catalogue, which includes more than 8000 galactic HII regions and HII region candidates, using a specific morphology in the mid-IR band as a selection criterion. We should also mention results of the Milky Way project \citep{2012MNRAS.424.2442S} and a catalogue by \cite{2017ApJ...846...64M}, based on observations from Spitzer and WISE space telescopes.

A number of detected objects increases each year, and finally we can lay out a solid basis for statistical and theoretical studies. A statistical analysis of HII regions \citep[see e.g.][]{2012A&A...542A..10A, 2013MNRAS.431.2006K,2014ApJS..212....1A,2017ApJ...846...64M,Topchieva} is a powerful tool to advance further interpretation of observational data and to relate them to results of numerical investigations. This is important as there are still some key questions, which lack definite answers. We mention briefly some of them.

It is still not clear how the object size is related to its age. Three varieties of HII regions are distinguished, namely:
\begin{enumerate}
\item ultracompact and hypercompact HII regions (size less than 0.1 pc, electron density $> 10^{4}$ cm$^{-3}$);

\item classic HII regions (size of the order of a few parsec, electron density $\sim10^{2}$ cm$^{-3}$);

\item giant HII regions (size of the order of 100 pc, density $< 30$ cm$^{-3}$).
\end{enumerate}

It is possible that they all represent different stages of a single process. Specifically, it has been suggested in \cite{2007ARA&A..45..481Z} that smaller and denser HII regions are young, while less dense and more extended HII regions are older. However, the exact evolutionary relationships between HII regions of various kinds are still unclear, and we cannot be certain if they exist at all.

The second problem, which is hard to solve, is the identification of a star that ionizes a given object. A standard suggestion, which is routinely adopted in various models, is the central location of the star \citep{1979A&A....77..165G,2004ApJ...608..282A,2011ApJ...732..100D,2013ARep...57..573P,2015MNRAS.449..440A,2017MNRAS.469..630A}. However, the ionizing star in an HII region may reside not only in its center, but also on the periphery and even beyond the object. The latter two morphologies are usually referred to as champaign flow and blister, respectively. Examples of such morphologies were found in Sh2-212 by \cite{2008ASPC..387..338D} and \cite{2008A&A...482..585D}, in Orion Nebulae by \cite{2000AJ....120..382O}, in several bipolar HII regions by \cite{2012A&A...546A..74D} and \cite{2015A&A...582A...1D}. It is also possible that some HII regions are excited by several OB stars, e.g. RCW79 \citep{2010A&A...510A..32M}.

To attack all these problems we need a self-consistent evolutionary model of HII regions, which is able to reproduce distributions of density, temperature, velocity, and molecular abundances simultaneously. On the other hand, we need thoroughly analyzed observational data, suitable for comparison with theoretical results. In this paper we present an analysis of photometry of HII regions in order to construct a large sample of objects, which can be used for comparison both with results of numerical simulations and with results from other studies of HII regions and complexes \citep{2012A&A...542A..10A, 2013MNRAS.431.2006K}. 

As an example of application of the presented photometric catalogue we estimate the PAH mass fraction ($q_{\rm PAH}$) and intensity of UV irradiation in the studied objects, using the grid of models by \cite{2007ApJ...657..810D}. Also we analyze possible differences in flux ratios between Galactic HII regions with resolved structure and extragalactic HII complexes, which are spatially unresolved.

In Section~2 we describe data processing. Section~3 contains photometric analysis of infrared ring nebulae images. In Section~4 results are presented and discussed.

\section{Data Processing}
\label{sec:data}

We use a catalogue presented in the work of \cite{Topchieva}. The 20cm New~GPS, created using the MAGPIS database of radio images of regions with Galactic coordinates $|b_{\rm gal}| < 0.8^{\circ}$ и $5^{\circ} < l_{\rm gal} < 48.5^{\circ}$, was used as the basis for this study. We identified compact sources of radio emission among the objects in this survey, toward which we performed a visual search of objects, which look like rings at 8\,$\mu$m and contain IR emission at 24\,$\mu$m and radio emission at 20 cm in their interiors. The catalogue is based on infrared survey data on 8 and 24\,$\mu$m, obtained with IRAC~\citep{2004ApJS..154...10F} and MIPS~\citep{2004ApJS..154...25R} instruments of the {\em Spitzer} space telescope. At longer wavelengths we used data from the {\em Herschel} Space Observatory science archive. Images on 70 and 160\,$\mu$m were obtained with the PACS instrument \citep{PACS}, while images on 250\,$\mu$m, 350\,$\mu$m, and 500\,$\mu$m were obtained with the SPIRE instrument \citep{SPIRE}. The total number of selected objects is 99. Four of them have not been identified in previous surveys of infrared ring nebulae.  Further they are designated as TWKK. Unlike other catalogues, this catalogue is specifically prepared as a resource for comparison with results of 1D spherically symmetric hydrodynamical computations. Thus, it only includes bright sources with a more or less regular structure.

Qualitatively we can assume that at 8 $\mu$m the major contribution to the object emission comes from infrared bands attributed to polycyclic aromatic hydro carbons (PAH). At 24\,$\mu$m the main sources of emission are presumable stochastically heated very small grains along with, probably, hot large grains \citep{2012ApJ...760..149P}. At longer wavelengths emission is mostly generated by colder large grains \citep{2011piim.book.....D}. At 250\,$\mu$m, 350\,$\mu$m, 500\,$\mu$m due to low angular resolution it is hard to distinguish between the object emission and the background (and foreground) emission \citep{2012A&A...537A...1A}, but we still consider these band in our study, trying to make the best effort in removing the background contribution.

\subsection{Aperture Photometry}\label{sec: flux}

Before the flux estimation, photometry data in all wavebands have been convolved to the same resolution using kernels from \cite{2011PASP..123.1218A}. We keep the original pixel size and take it into account, when computing fluxes.

\begin{figure}[h!]
\includegraphics[width=0.45\textwidth]{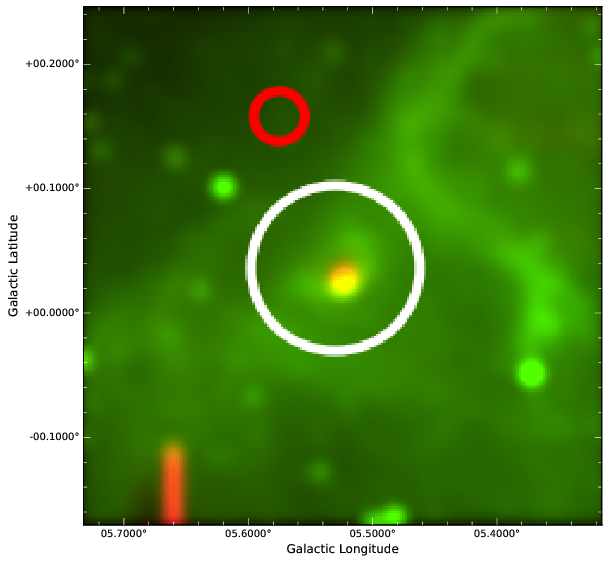}
\caption{Locations of the source and background apertures at 8\,$\mu$m map of the object CN67. A white circle shows the source aperture, while a red circle shows the aperture used for background estimation.}
\label{fig01}
\end{figure}

Size and location of a source aperture were selected using 8\,$\mu$m data on the base of the catalogue from \cite{Topchieva}. In order to estimate emission fluxes we need to get rid of background radiation and radiation from other sources, which are not related to the studied object (for example, stars and galactic background radiation). This is why we clean the image from point sources and subtract the background, estimated using a separate aperture, which is located in the darkest spot of the map beyond the object aperture. We believe that the background value measured at a brighter location or a background value, averaged over the entire image, can affect significantly the estimate of a useful source signal. Background aperture size depends on the extent of the area selected for background estimation (Fig. \ref{fig01}).

In all wavebands other than 8\,$\mu$m (24\,$\mu$m, 70\,$\mu$m, 160\,$\mu$m, 250\,$\mu$m, 350\,$\mu$m, and 500\,$\mu$m) for both source flux estimation and background estimation we use the same apertures as at 8\,$\mu$m. The custom-made Python scripts are utilized to compute total source fluxes in the above bands.

Apart from the HII region fluxes themselves we also consider flux ratios, or color indices, [$F_{24}$/$F_{8}$], [$F_{70}$/$F_{24}$], [$F_{160}$/$F_{24}$], [$F_{160}$/$F_{70}$] and compare them to the criteria suggested by \cite{2012A&A...537A...1A}. The authors have shown that these ratios can be used to discriminate between unresolved planetary nebulae and HII region. Below we check whether the criteria developed by \cite{2012A&A...537A...1A} can be applied to HII regions with spatially resolved structure. Also we relate our results in terms of flux ratios to those from the work of \cite{2013MNRAS.431.2006K}, where unresolved extragalactic HII regions have been considered.

\subsection{PAH mass fraction and UV field intensity}\label{sec: q}

Apart from simply computing flux ratios, we performed a somewhat more sophisticated analysis and estimated a PAH mass fraction, $q_{\rm PAH}$, using a grid of models from \cite{2007ApJ...657..810D}. The value of $q_{\rm PAH}$ is an important parameter in at least two respects. First, PAH emission is often considered as an indicator of the star formation rate, because corresponding transitions are excited by UV photons, which presumably trace the number of young stars. However, the relation between strength of the IR bands and the intensity of UV radiation can be non-trivial as UV photons both excite PAHs and destroy them. Thus, it is important to consider how $q_{\rm PAH}$ behaves on various spatial scales. Specifically, small value of $q_{\rm PAH}$ may indicate that organic dust particles are destroyed in HII regions \citep{2006A&A...446..877M,2007ApJ...665..390L}. Second, the evolution of PAHs in star-forming regions is important in the context of a general evolution of organic matter in the universe.

The grid of models from \cite{2007ApJ...657..810D} can be used to determine a UV field intensity $U$ in these regions. In this formalism, the radiation fields of the region under consideration is assumed to consist of two components, the minimum radiation field with intensity $U_{\min}$ and the enhanced radiation field with intensity $U$, distributed between $U_{\min}$ and some maximum value, $U_{\max}$. We consider a simplified situation of an isolated HII region, so we assume that there is a single value of radiation intensity, $U_{\min}$ (in other words, $U_{\max}=U_{\min}$), and denote it simply as $U$. This is also an important parameter as UV photons are crucial for thermal balance and dynamical evolution both inside the HII region and at its border, where a so-called photon-dominated region (PDR) is located. Both parameters are related to each other as UV photons both destroy PAH particles and excite them, so that they can generate IR emission bands. Also, in PDRs 0.1--1$\%$ of absorbed UV photons is transferred to suprathermal ($\sim$ 1 eV) photoelectrons, being ejected from dust and PAH particles. These electrons heat the gas, so $q_{\rm PAH}$ estimate is needed to evaluate the contribution of PAH to the thermal balance in these regions. In the grid of models by \cite{2007ApJ...657..810D} UV field intensity is measured in units of the intensity of the average radiation field in the Solar vicinity.

\section{Results}\label{sec:results}

\begin{figure}[h!]
\includegraphics[width=0.45\textwidth,clip=]{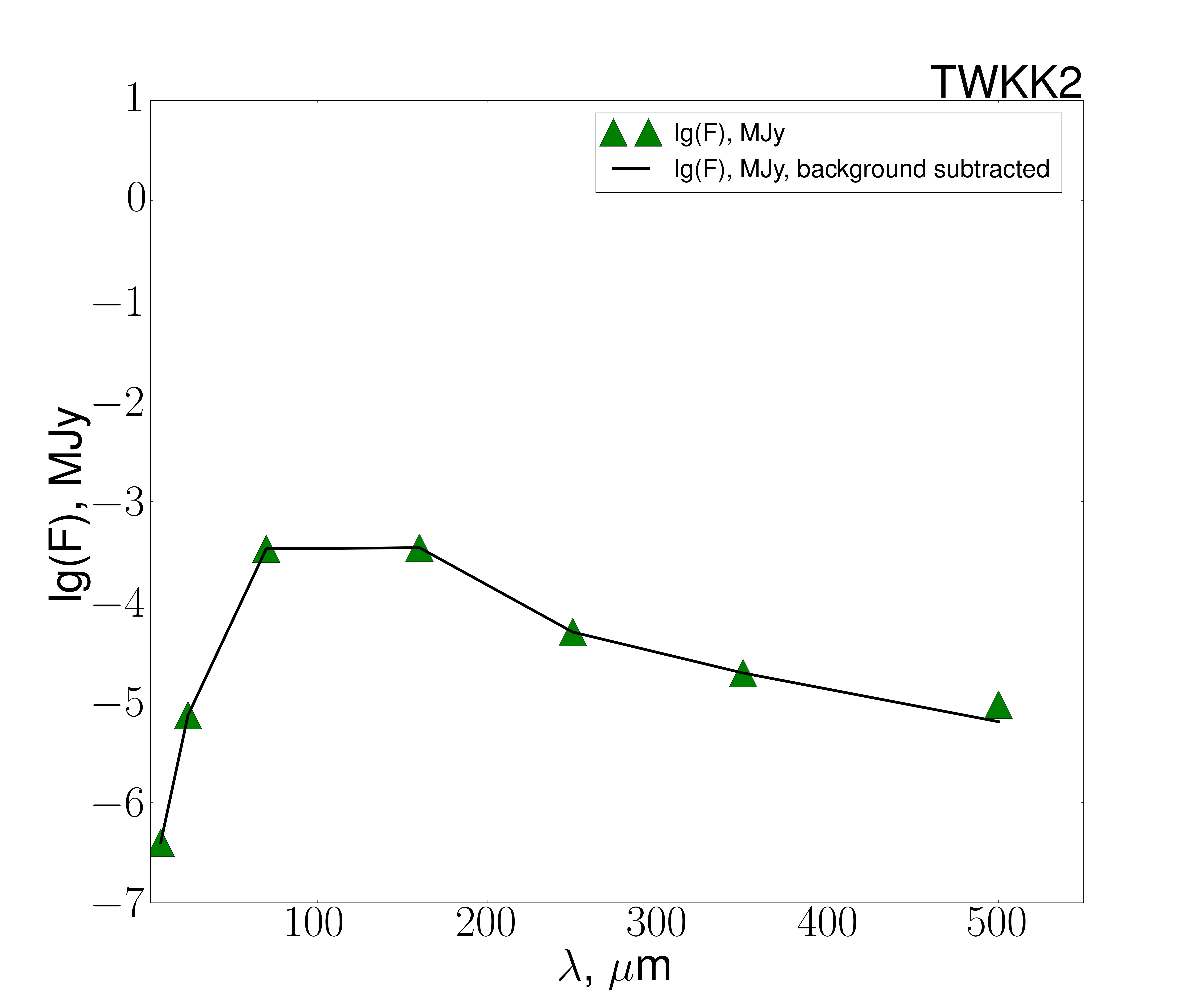}
\includegraphics[width=0.45\textwidth,clip=]{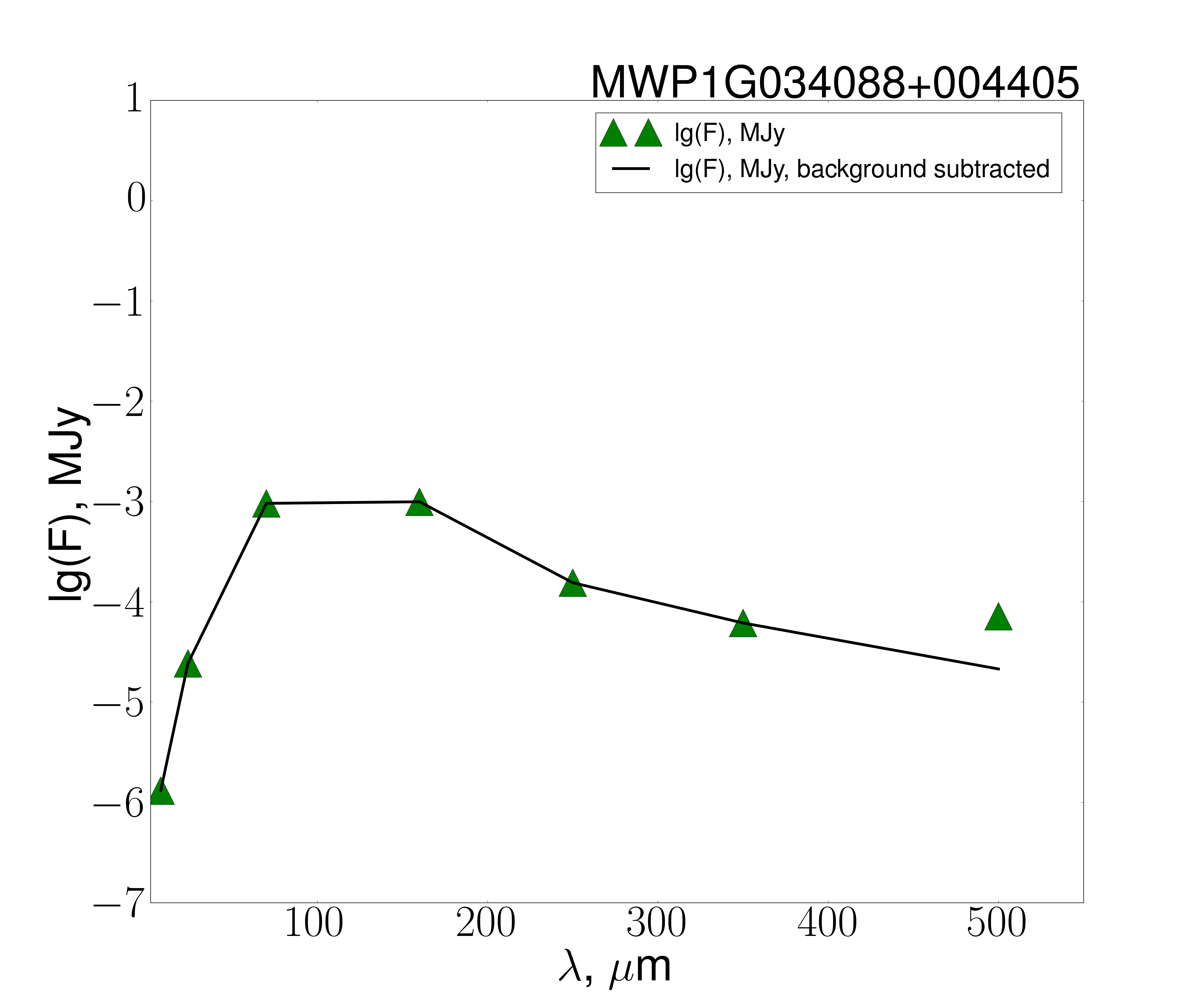}
\includegraphics[width=0.45\textwidth,clip=]{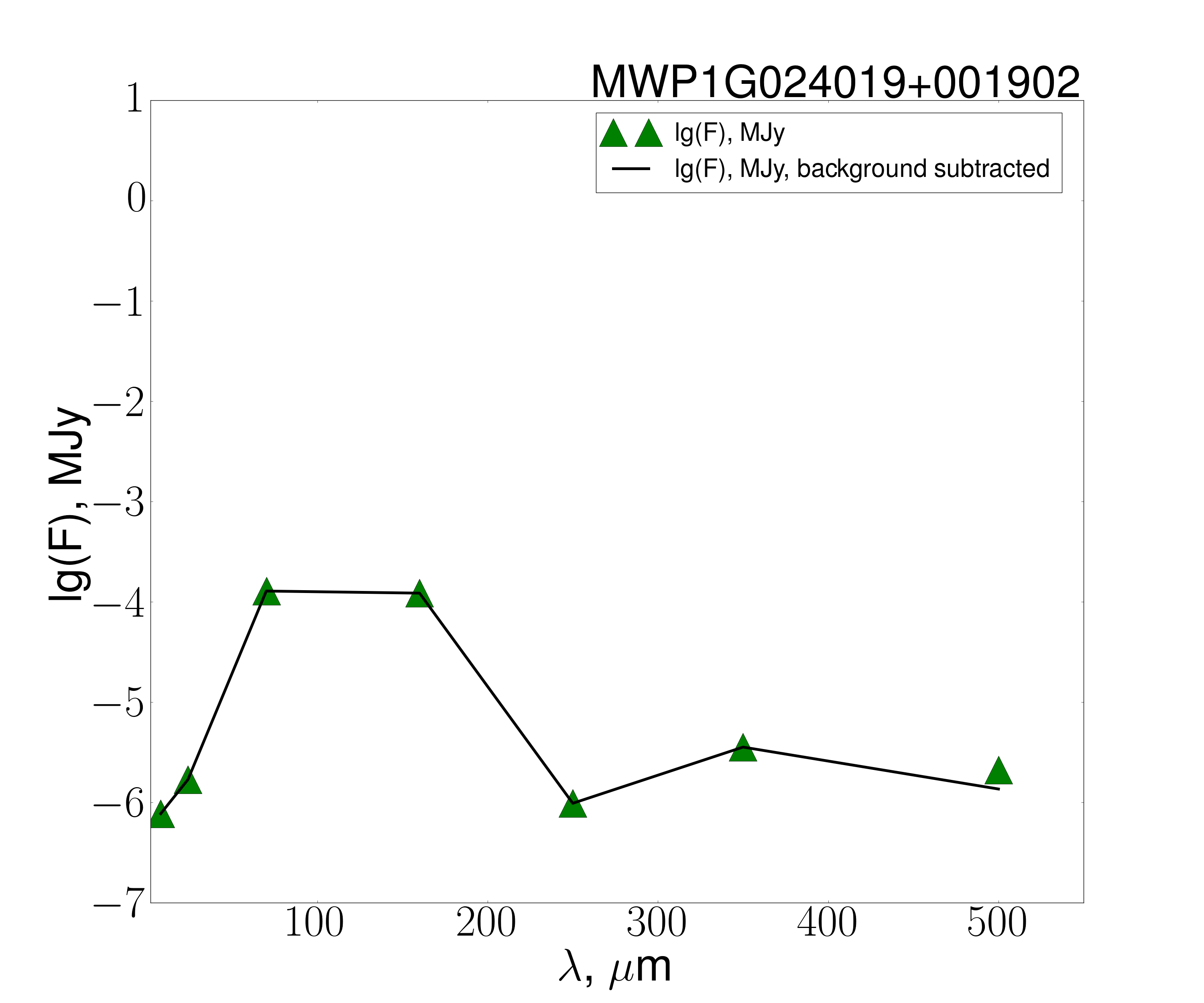}
\includegraphics[width=0.45\textwidth,clip=]{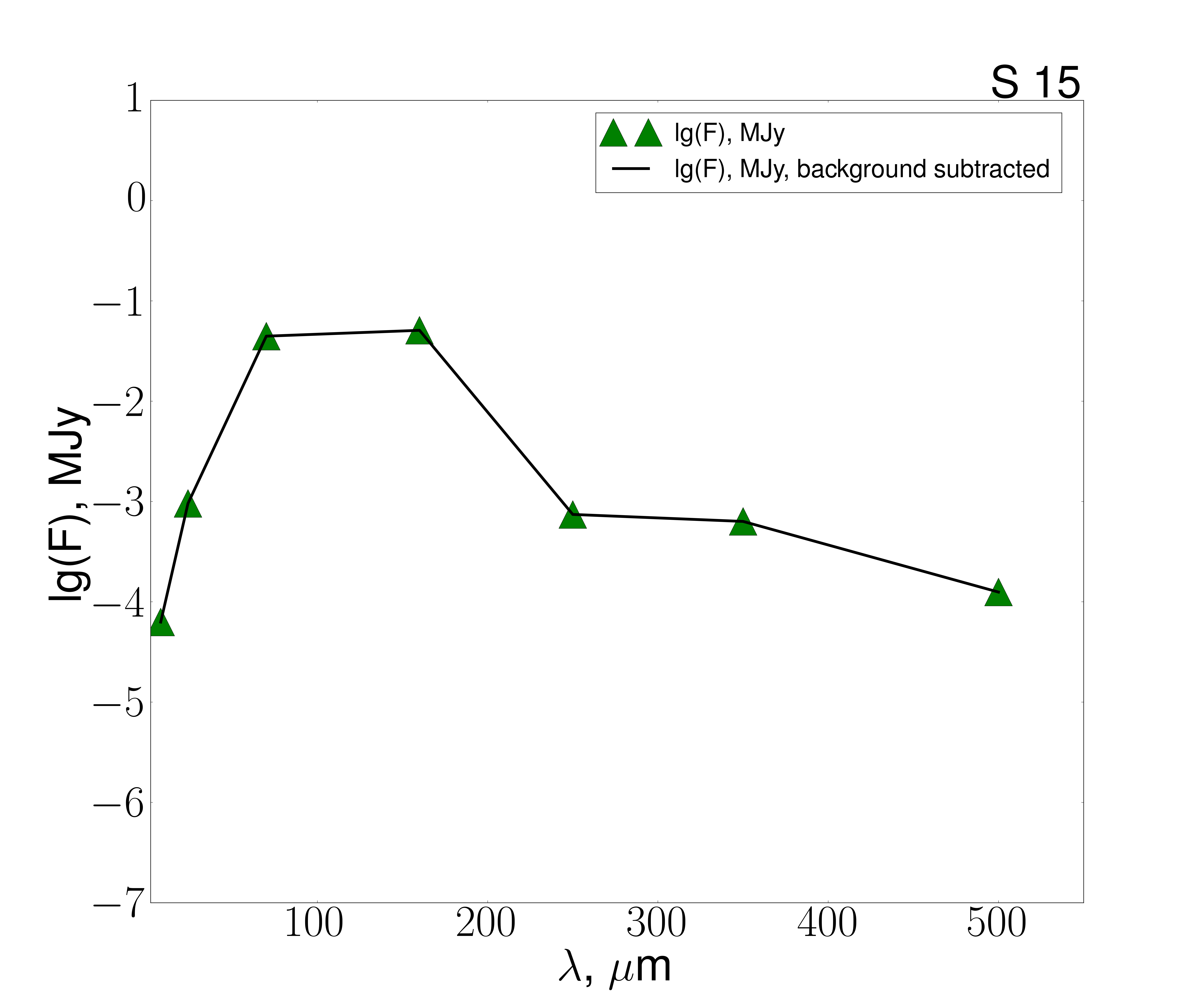}
\caption{Spectral energy distributions for objects TWKK2, MWP1G034088+004405, MWP1G024019+001902, and S15. Lines show the SED without background subtraction, while green triangles show flux values with background taken into account.}
\label{fig02}
\end{figure}

Performed flux measurements, presented in Table~\ref{tab:catal8mkm}, were used to construct spectral energy distributions (SED) for all the studied objects. We do see some variations in the SED shapes for different regions. Most objects have SEDs with the expected outline, that is, maximum emission at 70 and 160\,$\mu$m and shallow decrease toward 500\,$\mu$m. But there are some noticeable exceptions (Fig. \ref{fig02}). For example, objects TWKK2 and MWP1G034088+004405 show ``flat'' profiles, indicative of the excess presence of warmer dust. In objects MWP1G024019+001902 and S15 radiation flux decreases toward 250\,$\mu$m, which could mean incorrect background subtraction for these objects. We hope that HII region modeling will help to clarify these issues.

The second column of Table~\ref{tab:catal8mkm} contains the object designation. Galactic coordinates of the object are shown in columns 3 and 4. Fluxes in MJy are given in columns 5--11. Finally, in columns 12 and 13 we show estimates for PAH mass fraction and UV field intensity.

\begin{landscape}
\scriptsize 
\begin{longtable}{|p{0.5cm}|p{3cm}|p{1cm}|p{1cm}|p{1cm}|p{1cm}|p{1cm}|p{1.2cm}|p{1.2cm}|p{1.2cm}|p{1.2cm}|p{1cm}|p{1cm}|}
\caption{HII region parameters. The numbers between parentheses, $a(b)$, mean $a\times10^{b}$. Objects are taken from $^1$\cite{1994ApJS...91..347B}, $^2$\cite{2012MNRAS.424.2442S}, $^3$\cite{2006ApJ...649..759C}, $^4$new objects, $^5$\cite{2009A&A...507..795U}, $^6$\cite{2003yCat.5114....0E}.}\label{tab:catal8mkm}\\ \hline
No. & Object & $l_{\rm gal}$, $^{\circ}$ & $b_{\rm gal}$, $^{\circ}$ & $F_{8}$, MJy &  $F_{24}$, MJy &  $F_{70}$, MJy &  $F_{160}$, MJy &  $F_{250}$, MJy &  $F_{350}$, MJy &  $F_{500}$, MJy & $q_{\rm PAH}$ & $U$\\ \hline
\endfirsthead \hline
\multicolumn{13}{|c|}{\scriptsize\slshape(Continuation)} \\ \hline
$№$ & Object & $l_{\rm gal}$, $^{\circ}$ & $b_{\rm gal}$, $^{\circ}$ & $F_{8}$, MJy &  $F_{24}$, MJy &  $F_{70}$, MJy &  $F_{160}$, MJy &  $F_{250}$, MJy &  $F_{350}$, MJy &  $F_{500}$, MJy & $q_{\rm PAH}$ & $U$\\ \hline
\endhead \hline
\multicolumn{13}{|c|}{\scriptsize\slshape To be continued } \\ \hline
\endfoot \hline
\endlastfoot
\centering 
1  &  S15$^{3}$                    & 343.916    & --0.648      & 6.28(--5)   & 9.55(--4)   & 4.45(--2)   & 5.09(--2)   & 7.43(--4)   & 1.21(--3)   & 5.06(--4)     & 0.47         & 10.0\\
2  &  S21$^{3}$                    & 341.358    & --0.288      & 1.54(--5)   & 9.45(--6)   & 2.19(--3)   & 3.60(--3)   & 2.39(--4)   & 1.06(--4)   & 3.88(--5)     & 0.47         & 4.0\\
3  &  S44$^{3}$                    & 334.524    & 0.820        & 1.17(--4)   & 1.99(--4)   & 5.36(--2)   & 6.01(--2)   & 1.48(--3)   & 5.76(--4)   & 1.86(--4)     & 0.47         & 10.0\\
4  &  S123$^{3}$                   & 312.978    & --0.433      & 6.74(--5)   & 6.56(--5)   & 3.30(--2)   & 6.46(--2)   & 1.09(--3)   & 4.60(--4)   & 1.65(--4)     & 0.47         & 5.0\\
5  &  S145$^{3}$                   & 308.717    & 0.623        & 7.86(--4)   & 2.47(--3)   & 5.44(--1)   & 4.93(--1)   & 6.85(--3)   & 4.47(--3)   & 3.01(--3)     & 0.47         & 12.0\\
6  &  S167$^{3}$                   & 301.627    & --0.345      & 2.65(--5)   & 1.17(--4)   & 9.15(--3)   & 2.98(--2)   & 2.82(--3)   & 2.60(--3)   & 2.30(--3)     & 0.47         & 1.5\\
7  &  CN67$^{3}$                   & 5.526      & 0.037        & 1.18(--5)   & 1.65(--5)   & 7.88(--3)   & 1.35(--2)   & 2.33(--4)   & 9.81(--5)   & 3.52(--5)     & 0.47         & 4.0\\
8  &  CN77$^{3}$                   & 6.139      & --0.640      & 1.15(--5)   & 1.94(--4)   & 6.47(--3)   & 5.48(--3)   & 7.48(--4)   & 2.82(--4)   & 9.33(--5)     & 0.47         & 20.0\\
9  &  CN79$^{3}$                   & 6.202      & --0.334      & 2.94(--5)   & 5.89(--5)   & 9.41(--3)   & 1.40(--2)   & 9.77(--4)   & 4.49(--4)   & 1.60(--4)     & 0.47         & 7.0\\
10 &  CN111$^{3}$                  & 8.311      & --0.086      & 2.88(--5)   & 5.12(--5)   & 2.51(--2)   & 4.85(--2)   & 8.57(--4)   & 3.90(--4)   & 1.47(--4)     & 0.47         & 3.0\\
11 &  MWP1G008430--002800S$^{2}$   & 8.431      & --0.276      & 7.82(--6)   & 1.09(--5)   & 8.21(--3)   & 9.48(--3)   & 2.94(--4)   & 1.39(--4)   & 5.14(--5)     & 0.47         & 10.0\\
12 &  CN116$^{3}$                  & 8.476      & --0.277      & 7.63(--6)   & 9.18(--6)   & 6.70(--3)   & 7.88(--3)   & 1.33(--4)   & 5.83(--5)   & 2.14(--5)     & 0.47         & 10.0\\
13 &  N4$^{3}$                     & 11.893     & 0.747        & 2.28(--4)   & 2.47(--4)   & 8.18(--2)   & 8.82(--2)   & 1.25(--3)   & 9.02(--4)   & 6.43(--4)     & 0.47         & 10.0\\
14 &  MWP1G012590--000900S$^{2}$   & 12.595     & --0.090      & 2.26(--6)   & 3.78(--6)   & 3.54(--3)   & 4.33(--3)   & 6.43(--5)   & 2.50(--5)   & 8.20(--6)     & 0.47         & 10.0\\
15 &  MWP1G012630--000100S$^{2}$   & 12.633     & --0.017      & 7.63(--6)   & 1.15(--5)   & 1.23(--2)   & 2.18(--2)   & 4.59(--4)   & 2.24(--4)   & 8.45(--5)     & 0.47         & 3.0\\
16 &  N8$^{3}$                     & 12.805     & --0.312      & 2.43(--6)   & 3.35(--6)   & 1.19(--3)   & 2.12(--3)   & 1.71(--4)   & 8.00(--5)   & 3.97(--5)     & 0.47         & 3.0\\
17 &  MWP1G013213--001410$^{2}$    & 13.213     & --0.141      & 2.42(--5)   & 4.20(--5)   & 4.03(--2)   & 5.82(--2)   & 1.14(--3)   & 5.01(--4)   & 1.74(--4)     & 0.47         & 5.0\\
18 &  N13$^{3}$                    & 13.899     & --0.014      & 5.79(--6)   & 9.99(--6)   & 1.59(--3)   & 2.49(--3)   & 1.69(--4)   & 7.49(--5)   & 2.81(--5)     & 0.47         & 5.0\\
19 &  N14$^{3}$                    & 14.000     & --0.136      & 4.67(--4)   & 4.25(--4)   & 4.55(--2)   & 4.64(--2)   & 2.65(--3)   & 1.08(--3)   & 3.76(--4)     & 3.9-4.6      & 15.0\\
20 &  G014.175+0.024$^{6,1}$       & 14.175     & 0.022        & 2.06(--6)   & 4.72(--6)   & 2.81(--3)   & 3.53(--3)   & 5.56(--5)   & 2.27(--5)   & 7.51(--6)     & 0.47         & 10.0\\
21 &  MWP1G014210--001100S$^{2}$   & 14.206     & --0.110      & 2.03(--6)   & 4.33(--6)   & 1.70(--3)   & 1.37(--3)   & 3.90(--5)   & 1.57(--5)   & 5.20(--6)     & 0.47         & 25.0\\
22 &  MWP1G014390--000200S$^{2}$   & 14.388     & --0.024      & 9.87(--6)   & 8.41(--6)   & 4.68(--3)   & 2.81(--3)   & 1.09(--4)   & 4.31(--5)   & 1.46(--5)     & 0.47         & 25.0\\
23 &  MWP1G014480--000000S$^{2}$   & 14.490     & 0.022        & 4.85(--6)   & 7.33(--6)   & 6.54(--3)   & 1.15(--2)   & 2.20(--4)   & 1.01(--4)   & 3.62(--5)     & 0.47         & 4.0\\
24 &  MWP1G016390--001400S$^{2}$   & 16.391     & --0.138      & 1.50(--6)   & 2.63(--6)   & 1.93(--3)   & 2.43(--3)   & 3.96(--5)   & 1.63(--5)   & 5.53(--6)     & 0.47         & 8.0\\
25 &  MWP1G016429--001984$^{2}$    & 16.431     & --0.201      & 2.11(--5)   & 5.59(--5)   & 2.94(--2)   & 4.50(--2)   & 7.66(--4)   & 3.30(--4)   & 1.14(--4)     & 0.47         & 5.0\\
26 &  MWP1G016560+000056$^{2}$     & 16.560     & 0.002        & 2.11(--6)   & 1.59(--6)   & 1.83(--3)   & 3.96(--3)   & 6.95(--5)   & 2.96(--5)   & 1.05(--5)     & 0.47         & 2.5\\
27 &  MWP1G017626+000493$^{2}$     & 17.625     & 0.048        & 4.55(--6)   & 3.22(--6)   & 2.32(--3)   & 5.56(--3)   & 1.03(--4)   & 5.14(--5)   & 2.14(--5)     & 0.47         & 2.5\\
28 &  TWKK1$^{4}$                  & 17.805     & 0.074        & 1.80(--6)   & 9.74(--7)   & 6.62(--4)   & 9.98(--4)   & 1.24(--5)   & 8.90(--6)   & 3.27(--6)     & 0.47         & 4.0\\
29 &  N20$^{3}$                    & 17.918     & --0.687      & 7.81(--7)   & 1.38(--5)   & 5.94(--4)   & 9.54(--4)   & 2.17(--4)   & 1.06(--4)   & 3.85(--5)     & 0.47         & 4.0\\
30 &  MWP1G018440+000100S$^{2}$    & 18.442     & 0.013        & 6.87(--7)   & 1.68(--6)   & 1.43(--3)   & 1.68(--3)   & 3.23(--5)   & 1.67(--5)   & 6.68(--6)     & 0.47         & 10.0\\
31 &  MWP1G018580+003400S$^{2}$    & 18.582     & 0.345        & 2.24(--6)   & 7.81(--6)   & 1.66(--4)   & 1.95(--4)   & 1.42(--4)   & 6.68(--5)   & 2.57(--5)     & 4.60         & 12.0-15.0\\
32 &  N23$^{3}$                    & 18.679     & --0.237      & 9.81(--6)   & 1.98(--5)   & 9.11(--3)   & 9.03(--3)   & 2.95(--4)   & 1.42(--4)   & 5.46(--5)     & 0.47         & 12.0\\
33 &  MWP1G018743+002521$^{2}$     & 18.748     & 0.256        & 1.14(--5)   & 2.52(--5)   & 9.40(--3)   & 1.30(--2)   & 3.26(--5)   & 9.15(--5)   & 3.37(--5)     & 0.47         & 5.0-7.0\\
34 &  MWP1G020387--000156$^{2}$    & 20.388     & --0.017      & 6.80(--6)   & 4.68(--6)   & 4.76(--3)   & 7.25(--3)   & 9.21(--5)   & 8.67(--5)   & 3.71(--5)     & 0.47         & 5.0\\
35 &  MWP1G02100--000500S$^{2}$    & 21.005     & --0.054      & 1.53(--6)   & 3.45(--6)   & 2.51(--4)   & 4.13(--4)   & 7.00(--6)   & 2.98(--6)   & 1.08(--6)     & 0.47         & 3.0\\
36 &  N28$^{3}$                    & 21.351     & --0.137      & 2.56(--5)   & 2.46(--5)   & 3.46(--3)   & 3.76(--3)   & 2.18(--4)   & 1.88(--4)   & 7.00(--5)     & 0.47         & 10.0\\
37 &  N31$^{3}$                    & 23.842     & 0.098        & 5.24(--6)   & 7.14(--6)   & 6.40(--3)   & 8.22(--3)   & 1.33(--4)   & 5.80(--5)   & 2.10(--5)     & 0.47         & 7.0\\
38 &  MWP1G023849--001251$^{2}$    & 23.848     & --0.127      & 7.82(--6)   & 1.01(--5)   & 5.51(--3)   & 7.15(--3)   & 1.07(--4)   & 4.65(--5)   & 1.69(--5)     & 0.47         & 7.0\\
39 &  MWP1G023881--003497$^{2}$    & 23.881     & --0.350      & 4.37(--6)   & 8.28(--6)   & 5.50(--4)   & 5.14(--4)   & 4.68(--6)   & 7.61(--6)   & 2.61(--6)     & 0.47-1.2     & 12.0\\
40 &  N32$^{3}$                    & 23.904     & 0.070        & 5.70(--6)   & 1.35(--5)   & 1.16(--2)   & 1.56(--2)   & 2.65(--4)   & 1.24(--4)   & 4.78(--5)     & 0.47         & 5.0\\
41 &  MWP1G023982--001096$^{2}$    & 23.982     & --0.110      & 1.53(--6)   & 3.22(--6)   & 2.40(--4)   & 1.58(--4)   & 1.11(--5)   & 4.97(--6)   & 1.79(--6)     & 2.5-4.6      & 25.0\\
42 &  MWP1G024019+001902$^{2}$     & 24.043     & 0.204        & 7.72(--7)   & 1.69(--6)   & 1.28(--4)   & 1.22(--4)   & 9.80(--7)   & 3.56(--6)   & 1.36(--6)     & 0.47         & 15.0\\
43 &  MWP1G024149--000060$^{2}$    & 24.153     & --0.011      & 3.12(--6)   & 3.94(--6)   & 3.15(--3)   & 3.27(--3)   & 1.59(--4)   & 3.30(--5)   & 1.22(--5)     & 0.47         & 12.0-15.0\\
44 &  N33$^{3}$                    & 24.215     & --0.044      & 5.68(--6)   & 1.83(--5)   & 8.11(--3)   & 8.44(--3)   & 1.23(--4)   & 5.39(--5)   & 1.93(--5)     & 0.47         & 12.0-15.0\\
45 &  TWKK3$^{4}$                  & 24.424     & 0.220        & 1.76(--5)   & 3.86(--5)   & 1.70(--2)   & 2.37(--2)   & 5.00(--4)   & 2.47(--4)   & 9.05(--5)     & 0.47         & 7.0\\
46 &  TWKK2$^{4}$                  & 24.460     & 0.506        & 3.97(--7)   & 7.39(--6)   & 3.38(--4)   & 3.46(--4)   & 4.99(--5)   & 1.95(--5)   & 6.38(--6)     & 0.47         & 20.0\\
47 &  MWP1G024500--002400$^{2}$    & 24.502     & --0.237      & 1.24(--5)   & 2.00(--5)   & 7.66(--3)   & 8.62(--3)   & 1.62(--4)   & 6.69(--5)   & 6.15(--6)     & 0.47         & 15.0\\
48 &  MWP1G024558--001329$^{2}$    & 24.558     & --0.133      & 3.25(--5)   & 2.09(--5)   & 3.39(--3)   & 5.52(--3)   & 4.02(--4)   & 1.92(--4)   & 7.16(--5)     & 1.12         & 5.0\\
49 &  MWP1G024649--001131$^{2}$    & 24.651     & --0.078      & 2.93(--6)   & 2.81(--6)   & 2.08(--3)   & 1.60(--3)   & 1.78(--5)   & 7.85(--6)   & 3.12(--6)     & 0.47         & 20.0\\
50 &  MWP1G01024699--001486$^{2}$  & 24.700     & --0.148      & 1.56(--5)   & 2.66(--5)   & 1.14(--2)   & 9.54(--3)   & 1.13(--4)   & 3.37(--5)   & 1.03(--5)     & 0.47         & 25.0\\
51 &  MWP1G024731+001580$^{2}$     & 24.736     & 0.158        & 1.63(--5)   & 2.23(--5)   & 1.07(--2)   & 1.27(--2)   & 3.28(--4)   & 1.40(--4)   & 4.82(--5)     & 0.47         & 10.0\\
52 &  MWP1G024920+000800$^{2}$     & 24.922     & 0.078        & 3.32(--6)   & 8.33(--6)   & 7.03(--3)   & 1.13(--2)   & 2.10(--4)   & 9.74(--5)   & 3.50(--5)     & 0.47         & 5.0\\
53 &  MWP1G025155+000609$^{2}$     & 25.155     & 0.061        & 2.97(--5)   & 2.72(--5)   & 1.12(--2)   & 1.02(--2)   & 1.13(--4)   & 7.42(--5)   & 2.56(--5)     & 0.47         & 15.0\\
54 &  MWP1G025723+00058$^{2}$      & 25.724     & 0.058        & 1.16(--5)   & 1.78(--5)   & 9.62(--3)   & 9.15(--3)   & 1.15(--4)   & 4.41(--5)   & 1.48(--5)     & 0.47         & 15.0\\
55 &  MWP1G025730--000200S$^{2}$   & 25.726     & --0.027      & 3.02(--6)   & 2.88(--6)   & 1.97(--3)   & 6.83(--4)   & 4.05(--5)   & 1.54(--5)   & 4.39(--6)     & 0.47         & -----\\
56 &  N42$^{3}$                    & 26.329     & --0.071      & 1.46(--5)   & 1.54(--5)   & 1.11(--2)   & 1.57(--2)   & 2.46(--4)   & 1.04(--4)   & 3.58(--5)     & 0.47         & 5.0-7.0\\
57 &  N43$^{3}$                    & 26.595     & 0.095        & 1.39(--5)   & 2.06(--5)   & 1.09(--2)   & 1.44(--2)   & 2.37(--4)   & 1.04(--4)   & 3.73(--5)     & 0.47         & 7.0\\
58 &  MWP1G026720+001700S$^{2}$    & 26.722     & 0.173        & 4.88(--6)   & 5.00(--6)   & 3.51(--3)   & 4.13(--3)   & 4.12(--5)   & 2.97(--5)   & 1.03(--5)     & 0.47         & 10.0\\
59 &  G027.492+0.192$^{6}$         & 27.496     & 0.197        & 3.15(--5)   & 8.70(--5)   & 3.84(--2)   & 3.20(--2)   & 3.94(--4)   & 1.51(--4)   & 5.14(--5)     & 0.47         & 20.0\\
60 &  MWP1G02671+00300S$^{2}$      & 27.613     & 0.028        & 2.65(--6)   & 2.28(--6)   & 1.47(--3)   & 1.40(--3)   & 3.62(--5)   & 1.47(--5)   & 5.04(--6)     & 0.47-1.2     & 12.0-15.0\\
61 &  MWP1G027905--000079$^{2}$    & 27.904     & --0.009      & 9.35(--6)   & 1.11(--5)   & 7.77(--3)   & 9.53(--3)   & 1.45(--4)   & 5.78(--5)   & 1.92(--5)     & 0.47         & 7.0-10.0\\
62 &  G027.9334+00.2056$^{6,5}$    & 27.931     & 0.205        & 3.11(--6)   & 9.73(--6)   & 6.03(--3)   & 6.81(--3)   & 9.49(--5)   & 3.76(--5)   & 1.20(--5)     & 0.47         & 10.0\\
63 &  MWP1G027981+000753$^{2}$     & 27.981     & 0.073        & 1.96(--5)   & 2.50(--5)   & 4.36(--3)   & 4.41(--3)   & 2.41(--4)   & 1.67(--4)   & 5.74(--5)     & 0.47         & 10.0-12.0\\
64 &  MWP1G028160--000300S$^{2}$   & 28.160     & --0.046      & 8.14(--6)   & 4.16(--6)   & 4.97(--3)   & 7.72(--3)   & 1.14(--4)   & 4.28(--5)   & 1.38(--5)     & 0.47         & 5.0\\
65 &  N49$^{3}$                    & 28.827     & --0.229      & 9.90(--5)   & 1.86(--4)   & 1.94(--2)   & 3.42(--2)   & 1.89(--3)   & 9.77(--4)   & 3.46(--4)     & 0.47         & 3.0-4.0\\
66 &  MWP1G029136--001438$^{2}$    & 29.134     & --0.144      & 3.43(--6)   & 4.20(--6)   & 3.07(--3)   & 3.60(--3)   & 5.50(--5)   & 2.38(--5)   & 8.68(--6)     & 0.47         & 10.0\\
67 &  N51$^{3}$                    & 29.156     & --0.259      & 5.12(--5)   & 1.05(--4)   & 4.80(--3)   & 5.64(--3)   & 2.54(--4)   & 1.28(--4)   & 6.91(--5)     & 3.9-4.6      & 12.0\\
68 &  MWP1G030020--000400S$^{2}$   & 30.022     & --0.041      & 2.95(--5)   & 4.65(--5)   & 8.18(--3)   & 8.26(--3)   & 4.49(--4)   & 1.71(--4)   & 5.35(--5)     & 0.47         & 15.0\\
69 &  MWP1G030250+002413$^{2}$     & 30.251     & 0.240        & 1.31(--5)   & 1.29(--5)   & 1.30(--2)   & 1.48(--2)   & 1.82(--4)   & 7.61(--5)   & 2.67(--5)     & 0.47         & 10.0\\
70 &  MWP1G03080+001100S$^{2}$     & 30.378     & 0.111        & 7.89(--6)   & 1.25(--5)   & 7.89(--3)   & 8.59(--3)   & 1.25(--4)   & 5.08(--5)   & 1.76(--5)     & 0.47         & 10.0\\
71 &  MWP1G030381--001074$^{2}$    & 30.381     & --0.109      & 3.39(--5)   & 4.56(--5)   & 2.71(--2)   & 2.05(--2)   & 2.25(--4)   & 1.61(--4)   & 6.58(--5)     & 0.47         & 20.0\\
72 &  MWP1G031066+000485$^{2}$     & 31.071     & 0.049        & 4.63(--6)   & 1.01(--5)   & 6.27(--3)   & 3.76(--3)   & 3.36(--5)   & 1.83(--5)   & 2.27(--6)     & 0.47         & 25.0\\
73 &  MWP1G032057+000783$^{2}$     & 32.055     & 0.076        & 1.76(--5)   & 2.97(--5)   & 3.25(--2)   & 4.20(--2)   & 7.57(--4)   & 3.63(--4)   & 1.32(--4)     & 0.47         & 7.0-10.0\\
74 &  N55$^{3}$                    & 32.101     & 0.091        & 3.78(--5)   & 3.30(--5)   & 7.68(--3)   & 1.05(--2)   & 1.65(--4)   & 7.03(--5)   & 2.46(--5)     & 0.47         & 3.0-5.0\\
75 &  MWP1G032731+002120$^{2}$     & 32.730     & 0.212        & 7.42(--6)   & 7.64(--6)   & 4.72(--3)   & 9.30(--3)   & 1.60(--4)   & 7.30(--5)   & 2.76(--5)     & 0.47         & 3.0-4.0\\
76 &  N57$^{3}$                    & 32.761     & --0.149      & 3.41(--6)   & 2.25(--6)   & 2.14(--3)   & 3.73(--3)   & 5.51(--5)   & 2.19(--5)   & 7.68(--6)     & 0.47         & 3.0-4.0\\
77 &  N60$^{3}$                    & 33.815     & --0.149      & 9.26(--6)   & 1.14(--5)   & 7.95(--3)   & 8.24(--3)   & 1.13(--4)   & 5.26(--5)   & 3.45(--5)     & 0.47         & 10.0\\
78 &  MWP1G034088+004405$^{2}$     & 34.087     & 0.441        & 1.31(--6)   & 2.43(--5)   & 9.58(--4)   & 9.94(--4)   & 1.55(--4)   & 6.15(--5)   & 2.14(--5)     & 0.47         & 10.0-12.0\\
79 &  MWP1G034680+000600S$^{2}$    & 34.684     & 0.067        & 2.42(--6)   & 4.31(--6)   & 3.04(--3)   & 3.26(--3)   & 4.92(--5)   & 1.99(--5)   & 6.65(--6)     & 0.47         & 10.0-12.0\\
80 &  N67$^{3}$                    & 35.544     & 0.012        & 1.07(--5)   & 2.80(--5)   & 9.25(--3)   & 6.60(--3)   & 2.28(--4)   & 8.54(--5)   & 2.98(--5)     & 0.47         & 20.0\\
81 &  MWP1G037196--004296$^{2}$    & 37.195     & --0.429      & 6.24(--6)   & 1.15(--5)   & 1.50(--3)   & 1.59(--3)   & 2.28(--5)   & 2.17(--5)   & 8.09(--6)     & 0.47         & 10.0-12.0\\
82 &  MWP1G037261--000809$^{2}$    & 37.258     & --0.078      & 1.96(--5)   & 2.63(--5)   & 1.24(--2)   & 1.45(--2)   & 1.74(--4)   & 5.65(--5)   & 2.55(--5)     & 0.47         & 10.0\\
83 &  MWP1G037349+006876$^{2}$     & 37.351     & 0.688        & 1.10(--5)   & 2.11(--4)   & 8.58(--3)   & 9.27(--3)   & 1.27(--4)   & 5.17(--5)   & 1.93(--5)     & 0.47         & 10.0\\
84 &  N70$^{3}$                    & 37.750     & --0.113      & 5.75(--6)   & 2.46(--5)   & 1.05(--2)   & 1.21(--2)   & 1.73(--4)   & 7.28(--5)   & 2.52(--5)     & 0.47         & 10.0\\
85 &  G038.550+1648$^{6}$          & 38.551     & 0.162        & 2.69(--6)   & 9.07(--6)   & 3.84(--3)   & 4.05(--3)   & 6.18(--5)   & 2.43(--5)   & 8.43(--6)     & 0.47         & 10.0-12.0\\
86 &  N73$^{3}$                    & 38.736     & --0.140      & 2.10(--5)   & 1.31(--5)   & 3.68(--3)   & 7.53(--3)   & 5.28(--4)   & 2.23(--4)   & 7.92(--5)     & 0.47         & 3.0-4.0\\
87 &  N78$^{3}$                    & 41.228     & 0.169        & 1.92(--6)   & 1.78(--6)   & 1.13(--3)   & 1.70(--3)   & 2.67(--5)   & 1.11(--5)   & 4.27(--6)     & 0.47         & 5.0-7.0\\
88 &  G041.378+0.035$^{6,1}$       & 41.378     & 0.034        & 6.48(--6)   & 1.46(--5)   & 6.13(--3)   & 5.97(--3)   & 9.20(--5)   & 3.81(--5)   & 1.31(--5)     & 0.47         & 15.0-20.0\\
89 &  N79$^{3}$                    & 41.513     & 0.031        & 2.01(--5)   & 4.46(--5)   & 2.22(--2)   & 2.32(--2)   & 3.12(--4)   & 1.26(--4)   & 4.33(--5)     & 0.47         & 10.0-12.0\\
90 &  TWKK4$^{4}$                  & 41.595     & 0.160        & 1.42(--6)   & 1.89(--6)   & 9.89(--4)   & 1.10(--3)   & 1.58(--5)   & 6.42(--6)   & 2.18(--6)     & 0.47         & 10.0-12.0\\
91 &  N80$^{3}$                    & 41.932     & 0.033        & 1.42(--5)   & 1.55(--5)   & 2.12(--3)   & 3.62(--3)   & 2.64(--4)   & 1.31(--4)   & 5.57(--5)     & 0.47         & 3.0-4.0\\
92 &  N89$^{3}$                    & 43.739     & 0.114        & 5.14(--6)   & 8.00(--5)   & 3.07(--3)   & 3.19(--3)   & 4.45(--5)   & 2.72(--5)   & 1.70(--5)     & 0.47         & 10.0-12.0\\
93 &  N90$^{3}$                    & 43.774     & 0.060        & 2.39(--5)   & 2.05(--5)   & 7.07(--4)   & 2.30(--3)   & 3.15(--4)   & 1.06(--4)   & 4.39(--5)     & 0.47         & 1.0-1.5\\
94 &  MWP1G045540+000000S$^{2}$    & 45.544     & --0.005      & 5.72(--6)   & 1.33(--5)   & 5.87(--3)   & 4.48(--3)   & 5.86(--5)   & 2.35(--5)   & 7.87(--6)     & 0.47         & 20.0-25.0\\
95 &  N96$^{3}$                    & 46.949     & 0.371        & 2.96(--6)   & 6.78(--6)   & 6.97(--4)   & 7.35(--4)   & 3.89(--5)   & 2.77(--5)   & 1.13(--5)     & 0.47         & 10.0\\
96 &  N98$^{3}$                    & 47.027     & 0.218        & 3.21(--5)   & 3.16(--5)   & 4.58(--3)   & 9.43(--3)   & 7.49(--4)   & 3.54(--4)   & 1.33(--4)     & 0.47-1.2     & 3.0-4.0\\
97 &  MWP1G048422+001173$^{2}$     & 48.422     & 0.116        & 9.92(--6)   & 7.56(--6)   & 5.56(--3)   & 8.46(--3)   & 1.14(--4)   & 4.37(--5)   & 1.48(--5)     & 0.47         & 3.0-5.0\\
98 &  N102$^{3}$                   & 49.697     & --0.164      & 1.14(--5)   & 6.40(--5)   & 2.05(--2)   & 9.30(--3)   & 1.33(--4)   & 6.45(--5)   & 6.49(--5)     & 0.47         & 25.0\\
99 &  N121$^{3}$                   & 55.444     & 0.887        & 1.65(--6)   & 1.05(--5)   & 3.17(--4)   & 1.10(--3)   & 2.43(--5)   & 1.92(--5)   & 1.54(--5)     & 0.47         & 1.0-1.5\\
\hline

\end{longtable}
\end{landscape}

Analysis of the PAH mass fraction shows that almost all the objects, except for N14, N51, and MWP1G03080+001100S, are characterized by $q_{\rm PAH}$ less than 0.47$\%$. This is an upper limit as the grid of models by \cite{2007ApJ...657..810D} does not contain data for smaller values of $q_{\rm PAH}$. Nevertheless, this result agrees with our expectation that $q_{\rm PAH}$ in HII regions is smaller than the average value for the Milky Way galaxy (that is, about a few percent) due to PAH destruction in HII regions \citep{2006A&A...446..877M,2007ApJ...665..390L}. Note that there are some objects with the $q_{\rm PAH}$ value of about 4\% or greater. A case by case examination shows that these are regions with more complex morphology and with a significant unaccounted background contribution.

Values of UV field intensity $U$ are surprisingly low, from 1 to 10, which is lower than would be expected for the immediate vicinity of a massive star. However, this value is mostly determined by the far-IR part of the spectrum. In our objects this emission comes from the outer rings, which are located far away from the ionizing stars. A more detailed picture will arise, when we will analyze SED radial variations. Also, it would be interesting to relate $U$ value to the linear size of the object, checking a naive assumption that the object size can serve as a qualitative measure of its age. However, distances ought to be known for this, and kinematic distance estimates are only available for 11 objects from our sample.

It would be interesting to relate obtained values of $U$ and $q_{\rm PAH}$ to each other. Specifically, it has been shown in the work of \cite{2013MNRAS.431.2006K} that the PAH mass fraction tends to be smaller in star-forming complexes with greater $U_{\min}$. Obviously, we do not see a similar trend in our data. There are at least two reasons for this. First, we would like to emphasize that in most objects we were only able to get an upper limit for $q_{\rm PAH}$. Thus, strictly speaking, we cannot say for certain whether or not $q_{\rm PAH}$ (anti)correlates with $U$. It has been argued in \cite{2013MNRAS.431.2006K} that the ratio of fluxes at 8 and 24\,$\mu$m can be used as a substitute for $q_{\rm PAH}$ but this is only true when the fluxes are computed for large star-forming complexes. In the presented study this does not work as in our objects 8\,$\mu$m emission and 24\,$\mu$m emission are spatially separated from each other. Thus, we do not see any correlation between $U$ and $F_8/F_{24}$ either. Second, we interpret small upper limits for $q_{\rm PAH}$ as a signature of PAH destruction in the vicinity of an ionizing star or stars, so we may expect clearer relation between $q_{\rm PAH}$ and $U$ closer to the source of UV radiation. On the other hand, the derived $U$ value is to a large degree determined by far infrared emission, which mostly comes from the region periphery. This is why our derived $U$ values are close to those presented in \cite{2013MNRAS.431.2006K}, even though our $U$ is not an exact equivalent of $U_{\min}$ from \cite{2013MNRAS.431.2006K}. The latter value is more representative of the space between individual HII regions in a large star-forming complex.

In general, a major difference between the work of \cite{2013MNRAS.431.2006K} and the present study is in vastly different spatial scales. In most regions presented in \cite{2013MNRAS.431.2006K} $q_{\rm PAH}$ is about a few percent, which is significantly higher than in our regions. This implies that PAH destruction proceeds locally, in individual HII regions, while most emission at 8\,$\mu$m comes from more extended inter-region areas, which are not probed in the present study.

\begin{figure}[h!]
\includegraphics[width=0.45\textwidth]{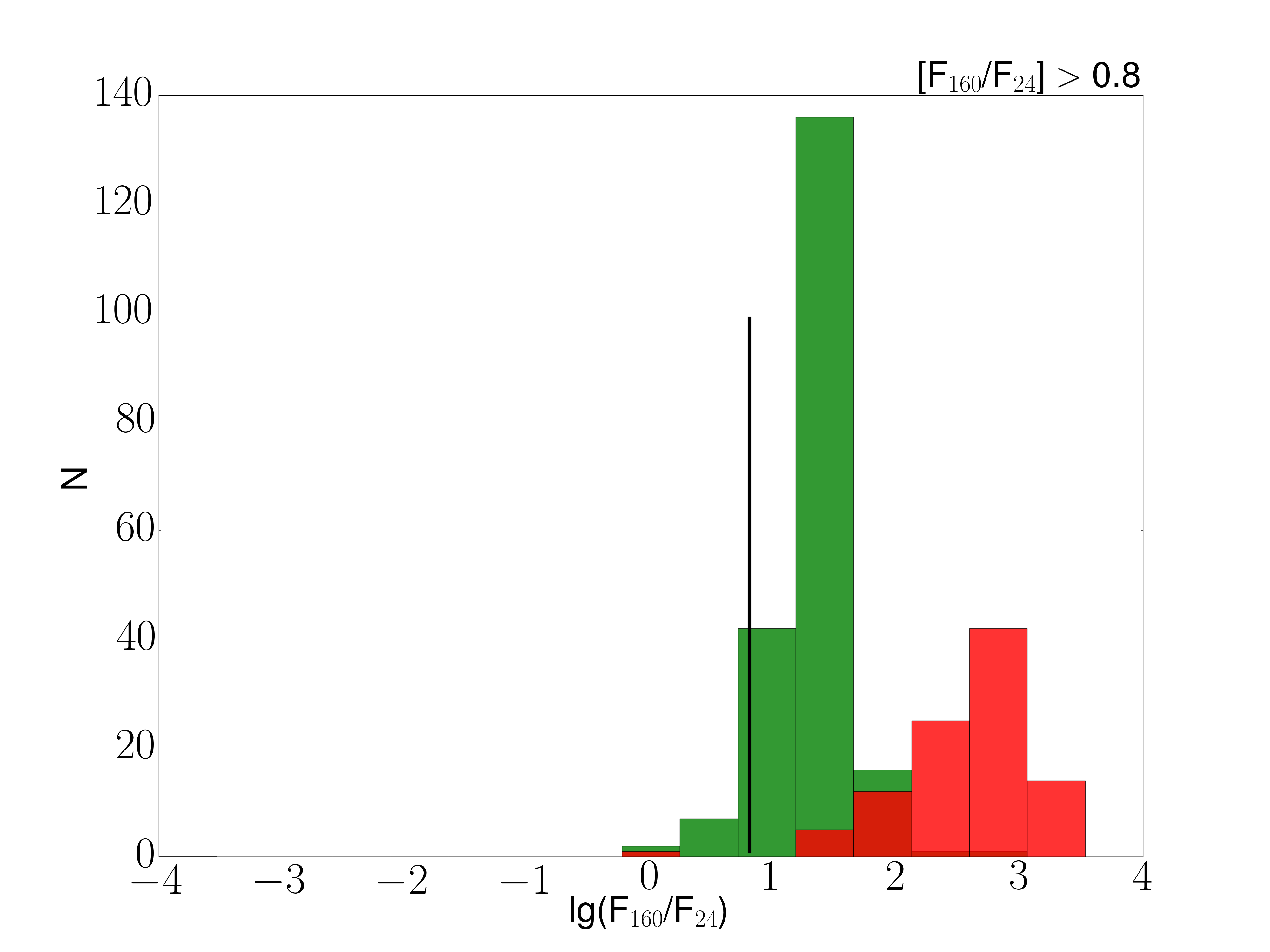}
\includegraphics[width=0.45\textwidth]{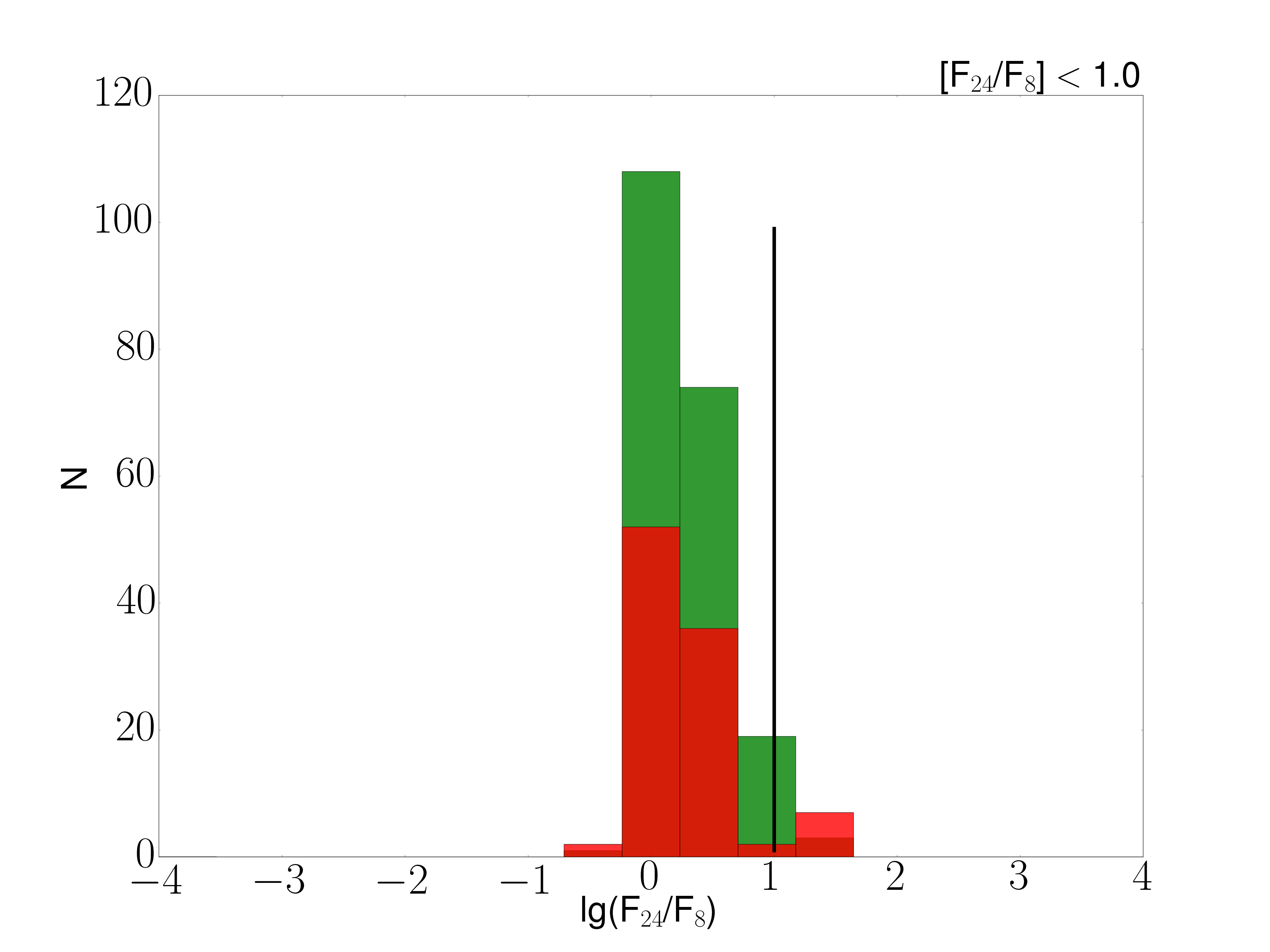}
\includegraphics[width=0.45\textwidth]{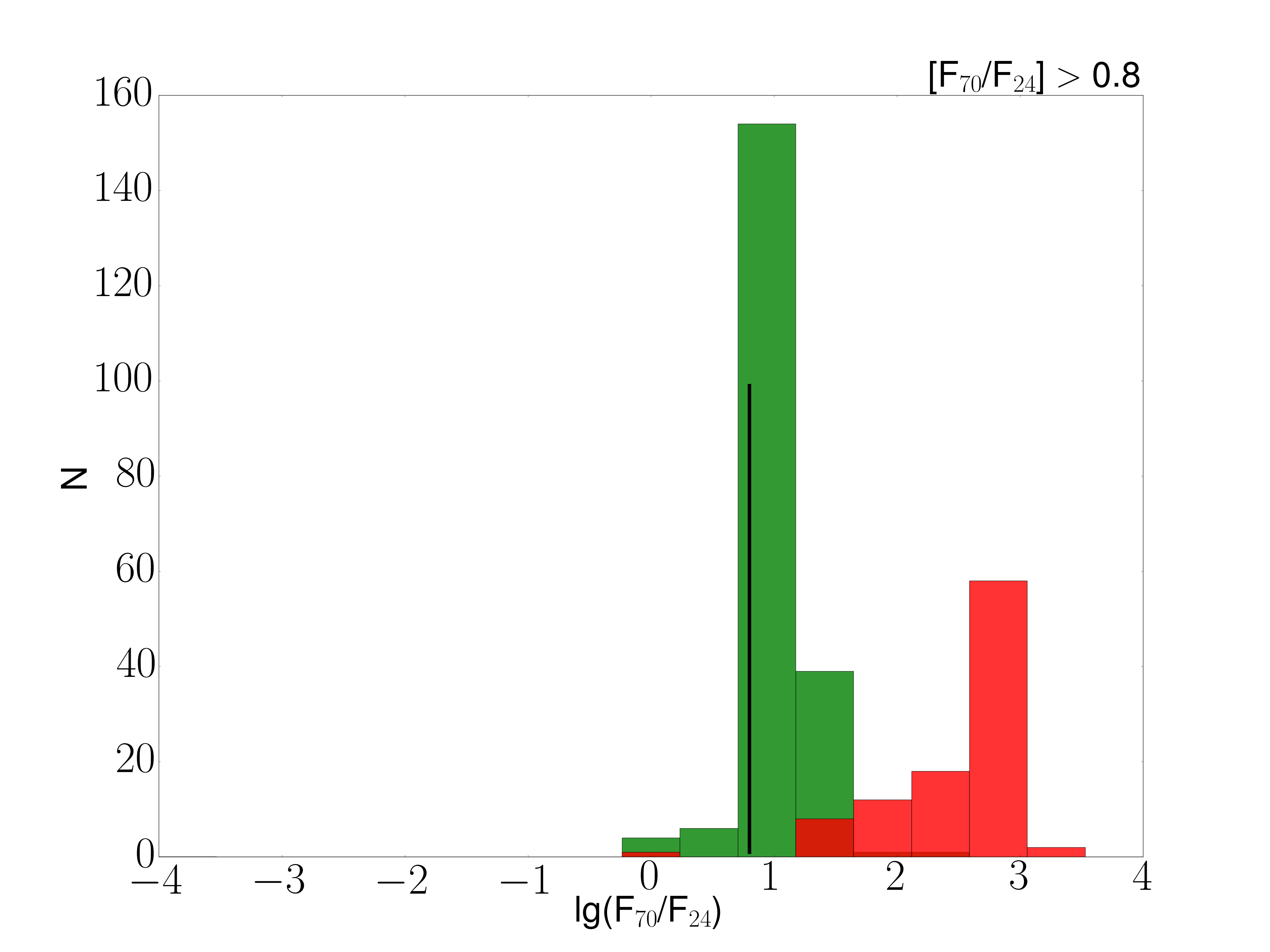}
\includegraphics[width=0.45\textwidth]{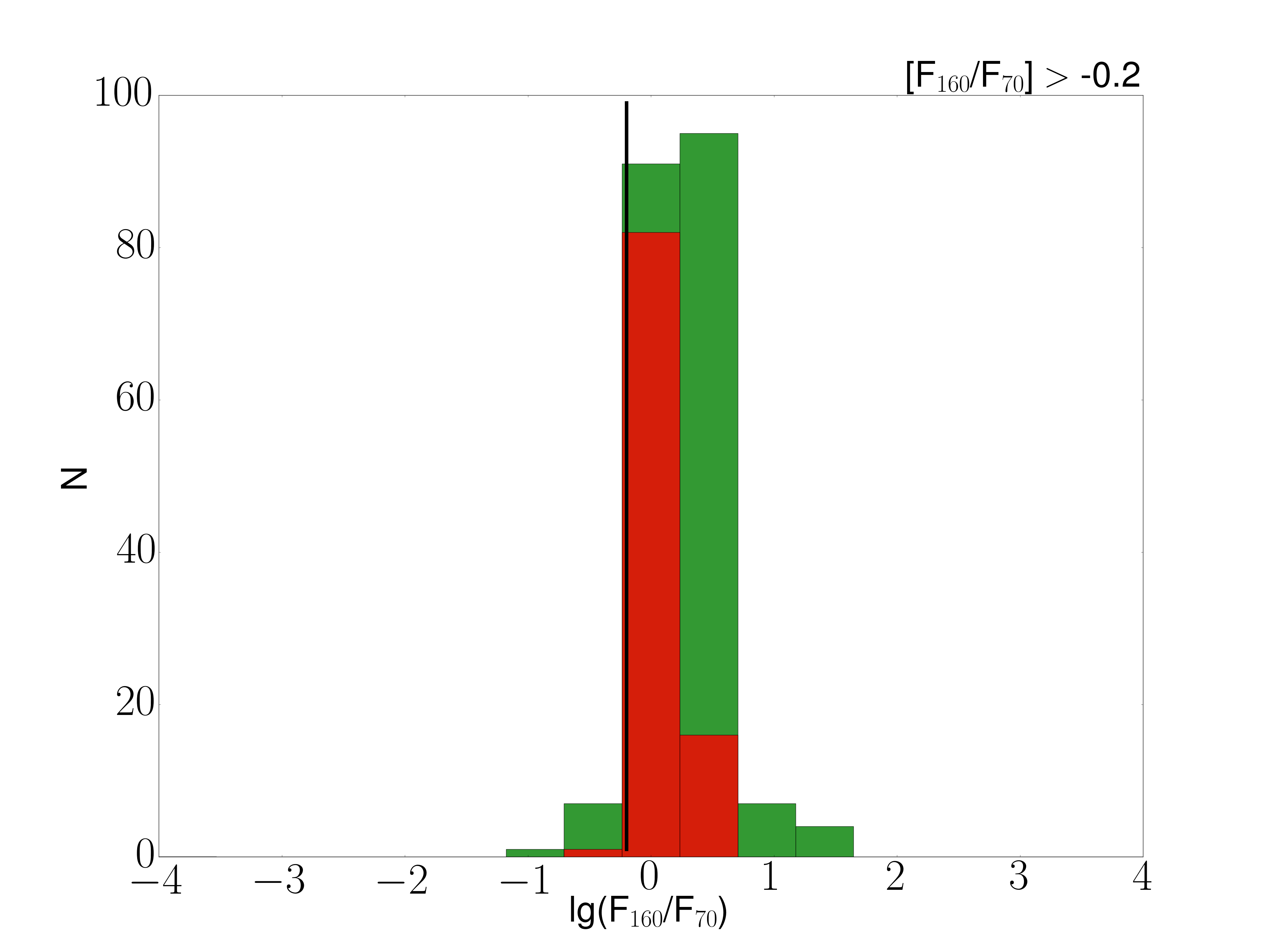}
\caption{Flux ratios for 99 HII regions studied in this paper (red bars) and for extragalactic HII complexes studied in \cite{2013MNRAS.431.2006K} (green bars). A black vertical line indicates a value discriminating between HII regions and planetary nebulae according to \cite{2012A&A...537A...1A}. The value is also shown in the top right corner of each panel. a) [$F_{24}/F_{8}$];  b) [$F_{70}/F_{24}$]; c) [$F_{160}/F_{24}$]; d) [$F_{160}/F_{70}$]. All the fluxes from this work are used with background subtraction.}
\label{fig03}
\end{figure}

In Fig.~\ref{fig03} we relate our flux ratios [$F_{24}$/$F_{8}$], [$F_{70}$/$F_{24}$], [$F_{160}$/$F_{24}$], and [$F_{160}$/$F_{70}$] to the criteria suggested in \cite{2012A&A...537A...1A} and similar ratios for extragalactic HII complexes presented by \cite{2013MNRAS.431.2006K} (square brackets indicate the logarithm of the corresponding ratio). In the top right corner of each panel we indicate the value of the flux ratio logarithm, which was indicated in \cite{2012A&A...537A...1A} as discriminating between HII regions and planetary nebulae. These values are also marked with vertical lines in each panel.

Obviously, most of our objects (red bars) and extragalactic HII regions (green bars) satisfy \cite{2012A&A...537A...1A} criterion and are indeed HII regions. However, there are few objects which are definitely HII regions, but would have been classified as planetary nebulae by \cite{2012A&A...537A...1A} constraints. Specifically, in some objects values of [$F_{24}$/$F_{8}$] ratios is greater than what is expected in HII regions, while for some objects the values of [$F_{70}$/$F_{24}$], [$F_{160}$/$F_{24}$], and [$F_{160}$/$F_{70}$] ratios are smaller than values expected in HII regions. This emphasizes that simple photometric criteria may fail, when they are applied to spatially resolved objects. We believe that the reason is two-fold. First, we draw the outer boundary of an object, relying on the location of the outer 8\,$\mu$m ring. However, 8\,$\mu$m emission allows drawing a boundary of an HII region only in the sense that it traces the location of a dense shell swept up by ionization and shock fronts. At the same time, in nearly all cases we also see a somewhat fainter 8\,$\mu$m emission, which extends well beyond the dense shell but is also related to the considered object. In other words, when we compare wide-scale maps of emission at 8 and 24\,$\mu$m, we see that 8\,$\mu$m emission is more extended than 24\,$\mu$m emission. Thus, we may inadvertently miss some 8\,$\mu$m emission which is located beyond the ring but still belongs to the object, while 24\,$\mu$m emission is accounted for entirely. On the other hand, due to lower angular resolution at longer wavelengths, some emission at these wavelengths may leak out of the 8\,$\mu$m-based aperture.

As for fluxes, computed by \cite{2013MNRAS.431.2006K} for extragalactic sources, their differences both from our data and from \cite{2012A&A...537A...1A} criteria are more significant, especially for [$F_{70}$/$F_{24}$] и [$F_{160}$/$F_{24}$] ratios. This is probably again related to a drastically different spatial scale probed in the work of \cite{2013MNRAS.431.2006K}. Star-forming complexes studied in that paper have typical linear sizes of a few hundred pc. In this case a single aperture includes both numerous individual HII regions and material between them. As we have mentioned above and as it had been found earlier in \cite{2008MNRAS.389..629B}, emission at 24\,$\mu$m is more compact than emission at other wavelength, while emission at the far-IR range is more diffuse. Thus, when an aperture corresponds to a large linear scale, we may expect a more significant contribution from emission at 70 and 160\,$\mu$m. This is why [$F_{70}$/$F_{24}$] и [$F_{160}$/$F_{24}$] flux ratios are greater in the work of \cite{2013MNRAS.431.2006K} than in the present paper.

\section{Conclusions}\label{conc}

The following results are presented in this work;
\begin{enumerate}
\item Total fluxes at 8\,$\mu$m, 24\,$\mu$m, 70\,$\mu$m, 160\,$\mu$m, 250\,$\mu$m, 350\,$\mu$m, and 500\,$\mu$m are estimated for 99 HII regions. This information can later be used for comparison with results of theoretical computations.

\item A PAH mass fraction $q_{\rm PAH}$ is estimated for these regions. In most regions we have only been able to obtain upper limits for $q_{\rm PAH}$, showing that the actual values are smaller than 0.47\%. This value is much lower than the average Galactic PAH mass fraction, which is about a few percent. We argue that this is a signature of local PAH destruction in HII regions.

\item Flux ratios [$F_{24}$/$F_{8}$], [$F_{70}$/$F_{24}$], [$F_{160}$/$F_{24}$], and [$F_{160}$/$F_{70}$] are estimated. It is shown that in some cases the criteria, suggested in \cite{2012A&A...537A...1A} to distinguish between HII regions and planetary nebulae, may fail when applied to spatially resolved objects.

\item Systemic differences with flux measurements in extragalactic HII complexes \citep{2013MNRAS.431.2006K} are caused by significantly different spatial scales.
\end{enumerate}

\normalem
\begin{acknowledgements}
This study is supported by the Program 7 of the Presidium of the RAS, ``Transitional and Explosive Processes in Astrophysics'', and the RFBR grant 17-02-00521. Astropy \citep{Astropy} package has been used to obtain presented results.
\end{acknowledgements}
  
\bibliographystyle{raa}
\bibliography{msRAA_2017_0251_R1}

\end{document}